\documentclass[X11names, dvipsnames,table]{article}
\usepackage{amsmath}
\usepackage[skip=0.2cm]{caption}
\usepackage{subcaption} 
\usepackage{fancyvrb}
\usepackage{url}
\usepackage{float}
\usepackage{multirow}
\usepackage{float}
\usepackage{verbatim}
\usepackage{musicography}
\usepackage{wasysym}
\usepackage{amsthm}
\usepackage{physics}
\usepackage{xcolor}
\usepackage[utf8]{inputenc}
\usepackage{authblk}
\usepackage{amsfonts}
\theoremstyle{definition}

\title{{\bf The Kolmogorov Complexity of  Irish traditional dance music}\\
}
\author[1]{Michael McGettrick\thanks{Michael.McGettrick@UniversityOfGalway.ie}}
\author[2]{Paul McGettrick\thanks{Paul.McGettrick@dkit.ie}}

\affil[1]{School of Mathematical and Statistical Sciences, University of Galway,
Galway H91 TK33, Ireland.}
\affil[2]{Department of Creative Arts,
Media and Music,
School of Informatics \& Creative Arts,
Dundalk Institute of Technology,
Dundalk A91 K584 Ireland.}

\date{\today}

\usepackage{graphicx}
\usepackage{geometry}
\begin{document}
\large
\maketitle

\begin{abstract}
We estimate the Kolmogorov complexity of 
melodies in Irish traditional dance music using
Lempel-Ziv compression. The ``tunes'' of 
the music are presented in so-called
``ABC notation'' as simply a sequence of 
letters from an alphabet: We have no 
rhythmic variation, with all notes being of 
equal length. Our estimation of algorithmic
complexity can be used to distinguish 
``simple'' or ``easy'' tunes (with 
more repetition) from ``difficult''
ones (with less repetition) which should
prove useful for students learning tunes.
We further present a comparison of two 
tune categories (reels and jigs) in 
terms of their complexity.
\end{abstract}

\section{Introduction}
The Kolmororov Complexity \cite{kolmogorov65, li_vitanyi_1997}
of s sequence of 
letters is defined to be the length of the 
shortest program which will output that 
sequence of letters. As such, it is known 
to be uncomputable. Nonetheless, it can 
be estimated in a reliable way using 
lossless compression techniques such as 
Lempel-Ziv compression \cite{DBLP:journals/tit/ZivL77}. Li and Sleep
\cite{DBLP:conf/ismir/LiS04, li2004melody} show how these 
ideas can be used to estimate the 
similarity between different melodies.

\section{Kolmogorov (algorithmic) complexity
via Lempel-Ziv (LZ) compression}
The Lempel-Ziv (family) of lossless compression
algorithms is known to be asymptotically optimal
\cite{286191}.
In the following sections,
we examine the compression of a common reel 
\emph{Sally Gardens} using two Lempel-Ziv variants: LZ77 and LZ78.
\subsection{LZ78 \cite{1055934} 
compression of the 
reel \emph{Sally Gardens}}

In Table \ref{tab1} we present the 
reel \emph{Sally Gardens} and in 
Table \ref{tab2} its compression
using LZ78. The compressed
version has 56 ``tokens''/``words''
with a compression ration of 
approximately 2.3.

\begin{table}[h]
    \centering
{\rowcolors{1}{green!60!yellow!30}{green!70!yellow!40}\begin{tabular}
{llll}
\hline
ggdg  bbgb &
DbEb  Dbab &
DDbD  EFGE &
Dbab  gede \\
\hline
ggdg  bbgb &
DbEb  Dbab &
DDbD  EFGE &
Dbab  gggg \\
\hline
DGGF GGDG &
GGBG AGFG &
EAAG AAEA &
AABG AGEG \\
\hline
DGGF GGDG &
GGBG AGFG &
DDbD EFGE &
Dbab gggg\\
\hline
\end{tabular}}
\caption{The reel
\emph{Sally Gardens} written in ABC notation.
For ease of reading, the tune is written
in 4 lines, but from an information theory
viewpoint, one ignores all spaces and new 
line characters and simply regards the 
tune as a sequence of characters/letters from
the alphabet \{a,b,c,d,e,f,g,A,B,C,D,E,F,G\}.
Here, lower/upper case letters are in the 
lower/upper octave respectively. We have taken 
standard musical liberty with the notation, 
representing all f\musSharp{}/F\musSharp{} as f/F respectively.}
\label{tab1}
\end{table}

\begin{table}[h]
\centering
{\rowcolors{1}{green!60!yellow!30}{green!70!yellow!40}\begin{tabular}%
{|r||l|l|l|l| l|l|l|l| l|l|l|l| l|l|l|l|}
\hline
& 1 & 2 & 3 & 4 &
5 & 6 & 7 & 8 &
9 & 10 & 11 & \fcolorbox{yellow}{yellow!30}{12} &
13 & 14 & 15 & 16 \\
\hline
\parbox{3cm}{\flushright Uncompressed string:}&g& gd& gb& b&
gbD& bE& bD& ba&
bDD& bDE& F& G&
E& D& bab&  ge\\
\hline
\parbox{3cm}{\flushright Compressed string:}&g& 1d& 1b& b& 
3D& 4E& 4D& 4a&
7D& 7E& F& G&
E& D& 8b& 1e\\
\hline
\parbox{3cm}{\flushright Cumulative note
count:}&1& 3& 5& 6& 
9& 11& 13& 15&
18& 21& 22& 23&
24& 25& 28& 30\\
\hline
\end{tabular}}{\tiny$\ \bullet\ \bullet$}

\vspace{0.5cm}

{\tiny$\bullet\ \bullet\ $}{\rowcolors{1}{green!60!yellow!30}{green!70!yellow!40}\begin{tabular}%
{|l|l|l|l| l|l|l|l| l|l|l|l| l|l|l|l|l|}
\hline
 17 & 18 & 19 & 20 &
21 & 22 & 23 & 24 &
25 & 26 & 27 & 28 &
29 & 30 & 31 & 32 & \fcolorbox{blue}{blue!30}{33}\\
\hline
d& e& gg& dg&
bb& gbDb& Eb& Db&
a& bDDb& DE& FG&
ED& babg&  ggg & DG & \fcolorbox{yellow}{yellow!30}{G}F\\
\hline
d& e& 1g& 17g& 
4b& 5b& 13b& 14b&
a& 9b& 14E& 11G&
13D& 15g& 19g & 14G & \fcolorbox{yellow}{yellow!30}{12}F\\
\hline
31& 32& 34& 36& 
38& 42 & 44& 46&
47& 51& 53& 55&
57& 61& 64& 66 & 68\\
\hline
\end{tabular}}{\tiny$\ \bullet\ \bullet$}

\vspace{0.5cm}

{\tiny$\bullet\ \bullet\ $}{\rowcolors{1}{green!60!yellow!30}{green!70!yellow!40}\begin{tabular}%
{|l|l|l| l|l|l|l| l|l|l|l| l|l|l|}
\hline
 34 & 35 & 36 &
    37 & \fcolorbox{red}{red!30}{38} & 39 & 40 &
41 & 42 & 43 & 44 &
45 & 46 & 47\\
\hline
 GG& DGG&
GB& GA& \fcolorbox{blue}{blue!30}{GF}G& EA&
A& GAA& EAA& AB&
GAG& EG& DGGF&  GGD\\
\hline
 12G& 32G& 
12B& 12A& \fcolorbox{blue}{blue!30}{33}G& 13A&
A& 37A& 39A& 40B&
37G& 13G& 35F& 34D\\
\hline
 70& 73& 
75& 77 & 80& 82&
83& 86& 89& 91&
94& 96& 100& 103\\
\hline
\end{tabular}}
{\tiny$\ \bullet\ \bullet$}

\vspace{0.5cm}

{\tiny$\bullet\ \bullet\ $}{\rowcolors{1}{green!60!yellow!30}{green!70!yellow!40}\begin{tabular}%
{|l|l|l|l| l|l|l|l| l|}
\hline
48& 49 & 50 & 51 & 52 &
53 & 54 & 55 & 56\\
\hline
GGG& DGA& \fcolorbox{red}{red!30}{GFG}D& DbD&
EF& GE& Dba& bg&
ggg\\
\hline
34G& 32A& \fcolorbox{red}{red!30}{38}D& 24D& 
13F& 12E& 24a& 4g&
19g\\
\hline
106& 109& 113& 116& 
118& 120 & 123& 125&
128\\
\hline
\end{tabular}}
\caption{ LZ78 compression of 
\emph{Sally Gardens} (see Table \protect\ref{tab1}). Each letter represents a quaver in the original melody. The 
\emph{compression ratio} is 
$128/56$ or approximately 2.3. For 
illustrative purposes, one particular
entry is highlighted in color.
Entry GFGD in column 50 is D appended
to GFG. In turn, GFG in column 38 is
G appended to GF. In turn GF in 
column 33 is F appended to G. Finally
G is in column 12.}
\label{tab2}
\end{table}




\subsection{LZ77 \cite{DBLP:journals/tit/ZivL77} compression of the 
reel \emph{Sally Gardens}}


\begin{table}[H]
\centering
{\rowcolors{1}{red!60!yellow!30}{red!70!yellow!40}\begin{tabular}%
{|r||l|l|l|l| l|l|l|l| l|l|l|l| l|l|l|l| l|l|}
\hline
& 1 & 2 & 3 & 4 &
5 & 6 & 7 & 8 &
9 & 10 & 11 & 12 &
13 & 14 & 15 & 16 & 17 & 18\\
\hline
\parbox{3cm}{\flushright Uncompressed string:}&g& g& d& g&
b& b& gb& D&
b& E& bDb& a&
bD& Db& D&  E & F & G\\
\hline
\parbox{3cm}{\flushright Compressed string:}&g& g& d& g& 
b& b& 4,2& D&
b& E& 8,3& a&
8,2& 9,2& D& E &F &G\\
\hline
\parbox{3cm}{\flushright Cumulative note
count:}&1& 2& 3& 4& 
5& 6& 8& 9&
10& 11& 14& 15&
17& 19& 20& 21& 22& 23\\
\hline
\end{tabular}}{\tiny$\ \bullet\ \bullet$}

\vspace{0.5cm}

{\tiny$\bullet\ \bullet\ $}{\rowcolors{1}{red!60!yellow!30}{red!70!yellow!40}\begin{tabular}%
{|l|l| l|l|l|l| l|l|l|l| l|l|l|}
\hline
19 & 20 &
21 & 22 & 23 & 24 &
25 & 26 & 27 & 28 &
29 & 30 & 31 \\
\hline
E& Dbab&
g& e& d& e&
ggdgbbgbDbEbDbabDDbDEFGEDbabg& gg& g& D&
G& G&  FG\\
\hline
E& 13,4& 
g& e& d& e&
1,29& 1,2& g& D&
G& G& 22,2\\
\hline
24& 28& 
29& 30 & 31& 32&
61& 63& 64& 65&
66& 67& 69\\
\hline
\end{tabular}}{\tiny$\ \bullet\ \bullet$}

\vspace{0.5cm}

{\tiny$\bullet\ \bullet\ $}{\rowcolors{1}{red!60!yellow!30}{red!70!yellow!40}\begin{tabular}%
{|l|l|l|l|l| l|l|l|l| l|l|l|l| l|l|}
\hline
32& 33 & 34 & 35 & 36 &
    37 & 38 & 39 & 40 &
41 & 42 & 43 & 44 &
45 & 46 \\
\hline
G& DGG& GDG& A&
GFG& E& A& AG&
AA& EAA& A& B&
GAG& E& GDGG\\
\hline
G& 65,3& 70,3& A&
67,3& E& A& 77,2&
82,2& 81,3& A& B&
76,3& E& 70,4\\
\hline
70& 73& 76& 77& 
80& 81 & 82& 84&
86& 89& 90& 91&
94& 95& 99\\
\hline
\end{tabular}}{\tiny$\ \bullet\ \bullet$}

\vspace{0.5cm}

{\tiny$\bullet\ \bullet\ $}{\rowcolors{1}{red!60!yellow!30}{red!70!yellow!40}\begin{tabular}%
{|l|l|}
\hline
47 & 48 \\
\hline
FGGDGGGDGAGFG & 
DDbDEFGEDbabgggg\\
\hline
68,13& 49,16\\
\hline
112& 128\\
\hline
\end{tabular}}
\caption{ LZ77 compression of 
\emph{Sally Gardens} (see Table \protect\ref{tab1}). Each letter represents a quaver in the original melody.
In the third row of the table, the notation
$i,j$ means one should go back to position 
$i$ in the
melody (position given in the fourth 
row of the table) and 
copy $j$ notes/letters.
The 
\emph{compression ratio} is 
$128/48$ or approximately 2.7. }
\label{tab3}
\end{table}

Comparing Tables \ref{tab1} and \ref{tab2} we
observe, from a musical viewpoint, clearly
repeated phrases are not picked up by 
the LZ78 compression. The second line in 
Table \ref{tab1} is in fact the same as the 
first line except for the last three notes 
`ggg'. 

In Table \ref{tab2}, columns 19 to 30,
we could have just issued an instruction to 
repeat columns 1 to 15 (with an added `g' at 
the end): It is clear better compression is 
possible.
\footnote{A fundamental limitation with
the LZ78 protocol is that the entries in the 
dictionary can only increase by one symbol
at a time.}

In LZ77, we do not use a dictionary, instead a sliding window is used, which will pick up (much) 
larger blocks of repeated letters/notes. In Table
\ref{tab3} we present the LZ77 compression of the 
same reel \emph{Sally Gardens}. Once can see 
immediately (columns 25, 47, 48)
the significantly larger blocks 
 of letters (musical phrases)
that are identified as occuring previously.

\section{Compression ratios for {\bf reels}}
\label{sec:reels}

We select a (musically) representative sample of 
60 reels from the online database 
\cite{theSess1}. The most common key for reels is D (about 45\%), the second most common G (about
30\%), with the remaining 25\% in a variety of 
keys. We examined tunes from across the key
spectrum. To enable fair comparison, we only examined ``standard'' length reels, i.e.\ ones
with exactly two parts: In our representation 
this corresponds to a sequence of 128 letters.
In summary,
\begin{itemize}
\item 
Each reel has two parts
\item 
Each part has exactly 64 (quaver) notes
\item 
Each bar has exactly 8 (quaver) notes
\item 
Any longer notes are written as sequences of
quavers adding up to the same note length.
So, any crotchet in the tune is written as 
two identical quavers, any dotted crotchet 
as three quavers, etc.
\item 
Each part of the tune has 8 bars, and overall the
tune has 16 bars
\end{itemize}

\begin{figure}[ht]
\captionsetup[subfigure]{aboveskip=-0.1cm,belowskip=0.7cm}
    \centering
    \begin{subfigure}[t]{\textwidth}
        \centering
\begin{Verbatim}
X: 1
T: Sally Gardens, The
Z: Jeremy
S: https://thesession.org/tunes/98#setting98
R: reel
M: 4/4
L: 1/8
K: Gmaj
|:G2GA BAGB|dBeB dBAB|d2Bd efge|dBAB GEDE|
GFGA BAGB|d2eB dBAB|d2Bd efge|dBAB G4:|
|:dggf g2de|g2bg ageg|eaag a2eg|a2bg ageg|
dggf g2de|g2bg ageg|d2Bd efge|dBAB G4:|

\end{Verbatim}
        \caption{The reel 
        \emph{Sally Gardens} in ``abc''
        notation, taken verbatim from \protect\cite{theSess1}}
    \end{subfigure}

    \begin{subfigure}[t]{\textwidth}
        \begin{center}
        \input{str4.txt}
        \end{center}
        \caption{Representation of the 
        tune using 128 quavers (colour/shading only for purposes of legibility)}
    \end{subfigure}

    \begin{subfigure}[t]{\textwidth}
        \begin{center}
        \input{str2.txt}
        \end{center}
        \caption{Compressed version: Contiguous letters are counted as one unit}
    \end{subfigure}

    \begin{subfigure}[t]{\textwidth}
        \begin{center}
        \input{str3.txt}
        \end{center}
        \caption{Compressed version: Every block of 2 or more letters is represented by {\tt [i,j]}, meaning ``go back to position 
        {\tt i} and copy 
        {\tt j} letters'' Note here, as is common in programming, we count from 0. This representation is of length 50.}
    \end{subfigure}
    \caption{Our python code processes the tune (initially in ``abc'' notation, (a)), though intermediate stages to reach 
 the compressed version (d)}\label{figH}
\end{figure}

\emph{Our analysis of the representative sample of 
60 reels gives an average compression ratio
of 2.79 (to two decimal places) with standard 
deviation 0.46.}

\subsection{Example: Comparison of 
reels with high and low complexity}

From the 60 tunes analyzed, the maximum
and minimum compression ratios obtained
were
\begin{itemize}
    \item 
    maximum compression ratio of 
    $\approx 4.92$ for \emph{The 
    Concertina Reel}
    \item minimum compression ratio of 
    exactly 2 for \emph{The Star of 
    Munster}
\end{itemize}
The details for this comparison are shown
in Table \ref{tabC}. The reader is encouraged to 
observe for themselves how much more 
repetitive \emph{The 
    Concertina Reel} is compared to 
    \emph{The Star of 
    Munster}.
\begin{table}[H]
    \centering
{\rowcolors{1}{red!60!blue!30}{red!70!blue!40}\begin{tabular}
{|l|l|l|c|}
\hline
Name &
Notes &
Compression &
Compressed length \\
\hline
\hline
\parbox{2.5cm}{\large The\\ Concertina Reel}
&\parbox{5cm}{\tt
AAFABAFA \textcolor{blue!80}{AAFABAFA} BBBABBBA \textcolor{blue!80}{BBBABAFA} AAFABAFA \textcolor{blue!80}{AAFABAFA} FABcdddA \textcolor{blue!80}{BAFEDDDD} AdddAddd \textcolor{blue!80}{AddABAFA} BBBABBBA \textcolor{blue!80}{BBBABAFA} AdddAddd \textcolor{blue!80}{AddABAFA} FABcdddA \textcolor{blue!80}{BAFEDDDD}} &
\parbox{5cm}{\tt
A A F A B [1, 3] [0, 8] B [16, 2] [15, 10] [5, 11] [0, 8] [2, 3] c d [52, 2] [3, 4] E D [60, 3] A [52, 4] [65, 6] [11, 22] [65, 15] [48, 16] }&
{\Large 26} \\
\hline
\parbox{2cm}{\large The Star of Munster}
&\parbox{5cm}{\tt
ccAcBBGB \textcolor{blue!80}{AGEFGEDG} EAABcBcd \textcolor{blue!80}{eaafgfed} cBAcBAGB \textcolor{blue!80}{AGEFGEDG} EAABcded \textcolor{blue!80}{cABGAAAA} eaabageg \textcolor{blue!80}{agbgagef} gfgagfef \textcolor{blue!80}{gfafgfdf} eaabageg \textcolor{blue!80}{agbgagef} gggeaaga \textcolor{blue!80}{bgafgeee}}&
\parbox{5cm}{\tt
c c A c B B G B A G E F [9, 2] D [9, 2] A A 
B [3, 2] c d e a a f g f e d [3, 2] [2, 3] 
[8, 2] [7, 14] [23, 2] [31, 2] [18, 2] G 
[17, 2] [17, 2] [24, 3] b a g e g [68, 2] b 
[71, 3] e [27, 3] [71, 3] [29, 2] [27, 3] [26, 4] d [29, 2] [65, 16] [112, 2] [24, 3] [71, 2] [74, 3] [27, 2] e [125, 2] }&
{\Large 64}\\
\hline
\end{tabular}}
\caption{Comparison of two reels with 
very low and very high 
Kolmogorov Complexity (shading/coloring only
for purposes of legibility)}
\label{tabC}
\end{table}

\section{Compression ratios for {\bf jigs}}
\label{sec:jigs}
From the same database 
\cite{theSess1}
we select a (musically) representative sample of 
60 jigs.
 We examined tunes from across the key
spectrum. To enable fair comparison, we only examined ``standard'' length jigs, i.e.\ ones
with exactly two parts: In our representation 
this corresponds to a sequence of 96 letters.
In summary,
\begin{itemize}
\item 
Each jig has two parts
\item 
Each part has exactly 48 (quaver) notes
\item 
Each bar has exactly 6 (quaver) notes
\item 
Any longer notes are written as sequences of
quavers adding up to the same note length.
So, any crotchet in the tune is written as 
two identical quavers, any dotted crotchet 
as three quavers, etc.
\item 
Each part of the tune has 8 bars, and overall the
tune has 16 bars
\end{itemize}
\emph{Our analysis of the representative sample of 
60 jigs gives an average compression ratio
of 2.61 (to two decimal places) with standard 
deviation 0.40.}

\section{Discussion of results}

\begin{figure}[ht]
    \centering
    \begin{subfigure}[t]{0.5\textwidth}
        \centering
        \includegraphics[height=2.4in]{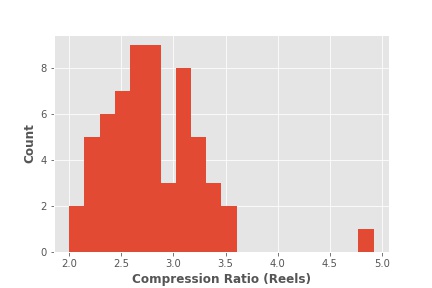}
        \caption{Distribution of compression
        ratios for 60 reels. The outlier with largest compression ratio ($\approx 4.92$) is \emph{The Concertina Reel}}
    \end{subfigure}%
    ~ 
    \begin{subfigure}[t]{0.5\textwidth}
        \centering
        \includegraphics[height=2.4in]{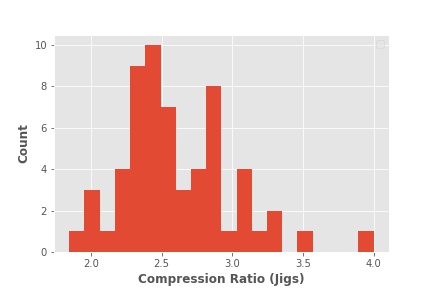}
        \caption{Distribution of compression
        ratios for 60 jigs.}
    \end{subfigure}
    \caption{Compression ratios for jigs and reels}\label{figI}
\end{figure}

In Figure \ref{figI} we display our results,
where we use 20 bins for the different
compression ratios.

In order to be able to account for 
differences in the (approximation to the)
Kolmogorov complexity of reels versus jigs,
we need to take in to account the different
lengths of these melodies / alphabet strings.
It should be clear that, for finite fixed 
alphabet size, the compression ratio will in general be larger for longer strings (for longer strings, letters will repeat more often).
In our case, the comparison is between 
strings of length 96 (jigs) and length
124 (reels). We would expect, all other things being equal, a slightly higher compression ration for the longer string (reels). To account for this effect (which 
is not intrinsic to the music), we generate 
sequences of random letters from the 13 
letter alphabet 
$A = \{a,b,c,d,e,f,g,h,i,j,k,l,m\}$. This 
corresponds to the lower octave, and most
of the higher octave in the tunes 
considered (the highest note in the high
octave is not used).

For strings of length 
$[50,100,150,200]$ we get average compression 
ratios of $[1.13, 1.24, 1.33, 1.40]$
respectively. In the region of interest
(lengths between 96 and 128), the 
dependence of compression ratio on 
string length is approximately linear.
In particular, for random strings of 
length 96/128 chosen from the alphabet
$A$, we get average compression ratios of 
1.23/1.29 respectively. We use this to 
normalize the results from Sections
\ref{sec:jigs} and \ref{sec:reels} to 
calculate what we would expect the 
average compression ratio for jigs to be,
were they of the same length as reels:
$\frac{1.29}{1.23}(2.61) 
\approx 2.73$.

Our result therefore is to conclude reels 
have slightly less intrinsic (Kolmogorov)
complexity than jigs, given the two 
averaged, normalized compression ratios of 
approximately 2.79 and 2.73 respectively.
Separately from comparing categories such as reels and jigs, our estimations of Kolmogorov complexity can be used to distinguish more repetitive tunes from less repetitive ones. This should be of use to those learning Irish traditional dance music, who may want to start with tunes of low Kolmogorov complexity.
We emphasize that we are \emph{not} concerned with efficiency issues in actually compressing the sequences considered, and in fact the compression
algorithms described may be very inefficient. We are only concerned with 
the size (length) of the resulting 
compression, how we can minimize that to 
get the best estimate of Kolmogorov 
complexity, and what that size can tell 
us about the tune being compressed.

In future work, we plan to apply this analysis
to other (large) families of tunes, such 
as hornpipes and polkas. We  are also 
exploring how Kolmogorov complexity, and 
 other techniques from information 
theory can be applied to harmonic music 
(for which there is a much larger corpus).

\bibliographystyle{plain}
\bibliography{references}
\end{document}